\documentclass[preprint]{iopart}

\usepackage{iopams}  
\usepackage{graphicx}
\usepackage{textcomp}
\usepackage[usenames]{color}

\definecolor{RED}{rgb}{1,0.1,0.1}
\definecolor{RED2}{rgb}{0.5,0,0}
\definecolor{GREEN}{rgb}{0,0.5,0}

\begin{document}

\title[XUV continua from relativistic plasma mirrors driven in the near-single-cycle limit]{Generation of XUV spectral continua from relativistic plasma mirrors driven in the near-single-cycle limit}

\author{Frederik B\"ohle$^{1}$, Maxence Th\'evenet$^{2}$, Ma\"imouna Bocoum$^{1}$, Aline Vernier$^{1}$,  Stefan Haessler$^{1}$ and Rodrigo Lopez-Martens$^{1}$}

\address{1 Laboratoire d'Optique Applique\'e, CNRS, Ecole Polytechnique, ENSTA Paris, Institut Polytechnique de Paris, 181 chemin de la Huni\`ere et des Joncherettes 91120 Palaiseau, France.}
\address{2 Lawrence Berkeley National Laboratory, Berkeley, California 94720, USA.}
\ead{stefan.haessler@ensta-paris.fr}

\begin{abstract}
We present experiments using relativistic-intensity 1.5-cycle laser fields at 1~kHz repetition rate to drive high-harmonic generation from surface plasmas with controlled density gradient. As a function of the CEP, we observe a transition from a modulated to a continuous SHHG spectrum, indicating the transition from double to high-contrast isolated attosecond pulse emission. Single shot-acquisitions of XUV spectral continua support isolated attosecond pulses with isolation degrees between 10 and 50 for the majority of the driving pulse CEPs, which represents a significant improvement in terms of isolation degree as well as repetition rate over previous results obtained with 2-cycle drivers at 10-Hz. 2D Particle-in-cell simulations corroborate this interpretation and predict percent-level efficiencies for the generation of an isolated attosecond pulse even without any spectral filtering.
\end{abstract}

%
%


\section{Introduction}
\label{sec:intro}

Surface high-harmonic generation (SHHG) from relativistic plasma mirrors has long been recognized as an efficient route to high-energy intense attosecond XUV pulses~\cite{tsakiris_route_2006,heissler_few-cycle_2012,jahn_towards_2019}. The phenomenon occurs when focusing an high-contrast visible to infrared driving laser in oblique incidence onto an optically polished solid target~\cite{thaury_high-order_2010}. Already the rising edge of a highly intense laser pulse creates a solid-density plasma on the target surface. If a sufficiently steep (scale length $L$ of a small fraction of the driving wavelength) plasma density gradient is retained, the incident light is nonlinearly reflected into a high-quality beam~\cite{dromey_diffraction-limited_2009,kahaly_direct_2013} with its spectrum extended into the XUV corresponding in the time domain to the emission of attosecond pulses. Plasma mirrors complement the established attosecond pullse generation method via high-harmonic generation (HHG) in gases at much lower intensities in the strong-field regime, since they exhibit no inherent limitation for the driving intensity. They thus allow exploiting fully ultra-high intensity lasers to convert an extremely large number of photons from a femtosecond laser into attosecond XUV pulses.  In strongly relativistic conditions, $a_0 \gg 1$, this is expected to occur with extremely high, percent-level conversion efficiencies~\cite{tsakiris_route_2006, anderBruegge2010nanobunching}. The laser-to-XUV conversion efficiencies of $\sim10^{-4}$ currently experimentally observed for plasma mirrors with $a_0\sim1$~\cite{Roedel2012ultrasteep,Heissler2014multimuJ,yeung_experimental_2017,jahn_towards_2019} already rival the highest demonstrated efficiencies of gas HHG~\cite{Takahashi2013gigawattatto,nayak_multiple_2018}. This makes SHHG on plasma mirrors one of the paramount candidates for greatly enhancing the available energy of attosecond XUV pulses. 

The interaction conditions on plasma mirrors may be roughly divided into a sub-relativistic and a relativistic regime, corresponding to a normalized vector potential $a_0= \sqrt{I [\mathrm{W cm}^{-2}] \,\lambda_0^2 [\mathrm{\mu m}^2]/(1.37\times10^{18})}<1$ and $> 1$, respectively, where $I$ is the laser intensity and $\lambda_0$ the central wavelength. These regimes are associated with two distinct SHHG mechanisms. In sub-relativistic conditions, coherent wake emission (CWE)~\cite{thaury_high-order_2010} dominates. This process requires an extremely steep plasma-vacuum interface ($L\sim\lambda/100$) and typically generates a spectral plateau extending up to the maximum plasma frequency given by the density of the solid target, which corresponds to $\approx30\:$eV photon energy for an SiO$_2$ target. 

For relativistic driving intensities, the dominating mechanism is described by a three-step push--pull--emission process~\cite{gonoskov_theory_2018,thevenet_physics_2016} repeating once per driving laser period. \textit{(i) ``Push'':} The incident laser field first pushes electrons through the magnetic force into the plasma, piling up a dense electron bunch and creating a corresponding restoring internal plasma field. \textit{(ii) ``Pull'':} As the laser field changes sign, the combined plasma and laser fields accelerate the electron bunch to a relativistic velocity towards the vacuum. \textit{(iii) ``Emission'':} High-harmonic emission is then described either as a pure phase modulation of the incident laser field through a temporal compression upon reflection on the relativistically moving critical-density plasma surface (``relativistic oscillating mirror'')~\cite{lichters_short-pulse_1996,gordienko_relativistic_2004,tsakiris_route_2006}, or as a coherent synchrontron emission (CSE) from the relativistically moving dense electron bunch as it gets accelerated orthogonally by the laser electric field~\cite{anderBruegge2010nanobunching, mikhailova_isolated_2012}. The generated spectra have a typical power-law decay $\propto\omega^{-p}$, with $p\approx1.3$--$3$ depending on the conditions, and for sufficiently high driving intensity can extend well beyond the spectral cutoff of the CWE mechanism. The optimal plasma density gradient scale length for this process is found to be $L\approx\lambda/10$~\cite{Roedel2012ultrasteep,kahaly_direct_2013,gao_double_2019}.

Temporally gating this process to a single dominant driving laser period and thus limiting SHHG to the emission of a single attosecond pulse remains extremely challenging with ultra-high-intensity laser systems, i.e. of terawatt to petawatt class with extreme temporal contrast of $\geq10^{10}$, since the known methods from gas HHG~\cite{calegari_temporal_2012} are not easily transferred. Proof-of principle demonstrations of polarization-gating~\cite{baeva_relativistic_2006,rykovanov_intense_2008,yeung_noncollinear_2015,chen_isolated_2018} and the attosecond lighthouse method~\cite{vincenti_attosecond_2012,wheeler_attosecond_2012} have been reported, but have the drawback of being rather costly in terms of achievable driving intensity.

More efficient but technically very challenging is intensity-gating through reduction of the laser pulse duration. For this approach, 2-cycle pulses with $\sim10\:$mJ energy have been generated by optical parametric chirped pulse amplifier (OPCPA) chains with $10\:$Hz repetion rate~\cite{rivas_next_2017,kessel_relativistic_2018}. These have been applied to generate SHHG with spectrally overlapping harmonics~\cite{jahn_towards_2019,kormin_spectral_2018} -- the essential prerequisite for an isolated attosecond pulse. Simulations supporting theses experiments~\cite{jahn_towards_2019} or a spectral interferometry analysis of SHHG spectra presenting a beating pattern of three unevenly spaced attosecond pulses~\cite{kormin_spectral_2018} have lead to the conclusion that for the optimal CEP value, an isolated attosecond pulse with isolation degree $\approx25$ (defined as main-to-satellite pulse temporal intensity ratio) can be generated after spectral selection of harmonics $>40\:$eV. The vast majority of driving pulse CEPs did however lead to emission of a double or triple attosecond pulse train. 

Our approach to intensity gating is based on the spatio-temporal confinement of few-mJ-energy pulses to a focal volume comparable to the laser wavelength cubed~\cite{albert_generation_2000}, which allows achieving relativistic interaction conditions at kHz-repetition rate~\cite{guenot_relativistic_2017,haessler_relativisticSHHG_2020}. To this end, we have developed a 1-kHz repetition rate laser that routinely delivers 3.5-fs (1.5-cycles of the 720-nm central wavelength) pulses with 1~TW peak power, based on a high-contrast Ti:sapphire double-CPA system followed by a power-scaled stretched-hollow-core-fiber compressor~\cite{jullien_doubleCPA_2014,bohle_compression_2014, ouille_SN2laser_2020}.  Here we report on SHHG from relativistic plasma mirrors driven by this 1.5-cycle laser, leading to XUV continua (10--25~eV) with a degree of residual spectral modulation varying as a function of the laser CEP and supporting an isolated attosecond pulse with isolation degree $\gtrsim25$ for half of the $2\pi$-CEP-range. We also predict theoretically that such a driver pulse should allow generating an attosecond pulse with isolation degree of 10 without any spectral filtering.

\section{Experiments}
\label{sec:exp}

\begin{figure}[tb]
	\centering
	\includegraphics[width=0.8\textwidth]{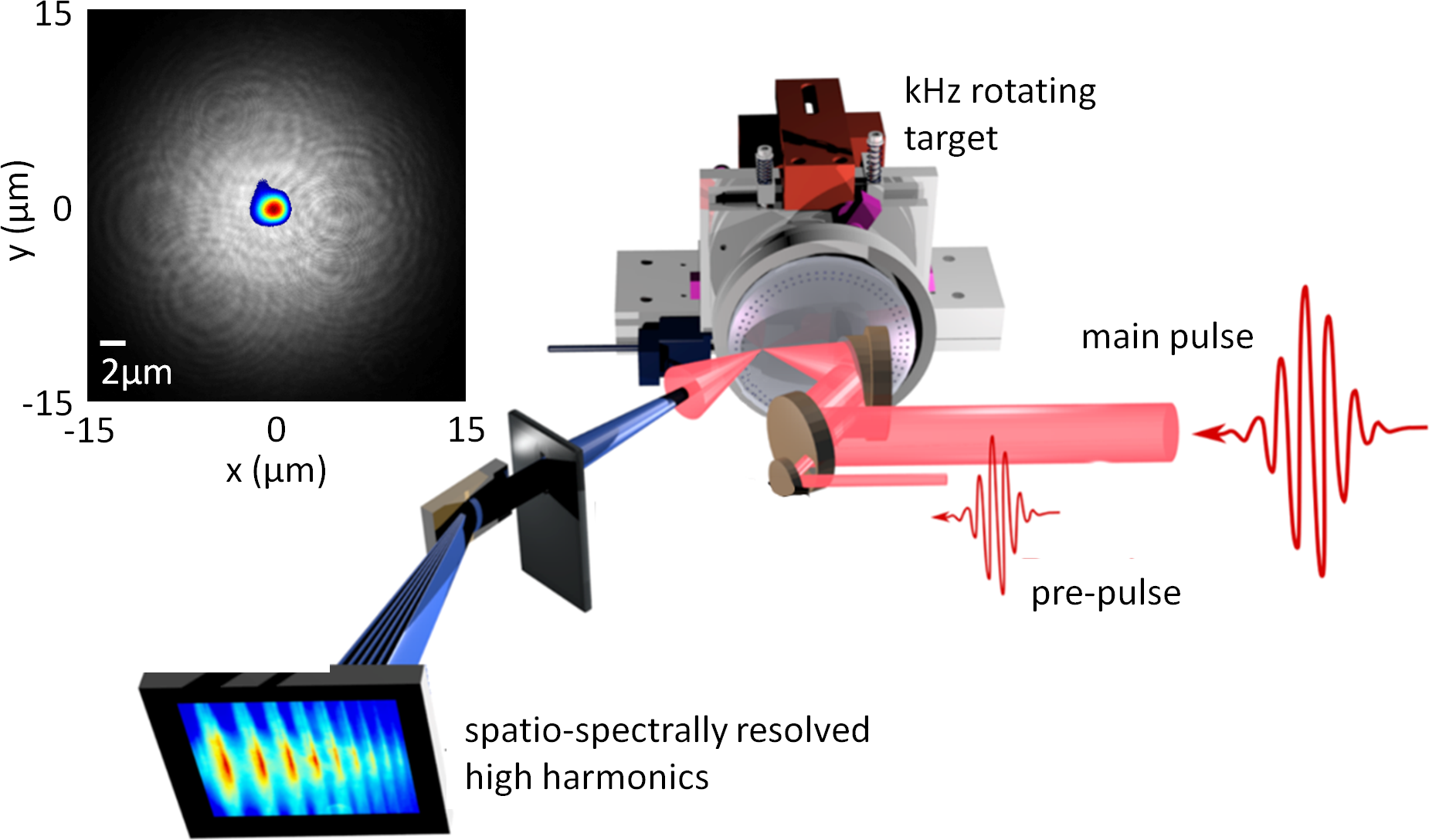}
	\caption{Schematics of the experimental setup for SHHG on a plasma mirror. The inset shows the on-target intensity profiles of the pre-pulse (in greyscale) and the main-pulse (in colorscale).}
	\label{fig:setup}
\end{figure}

For SHHG, a vacuum beamline sends the 3.5-fs laser pulses onto a rotating optically flat solid target with a controlled plasma density gradient. As illustrated in figure~\ref{fig:setup}, the p-polarized laser is focused using an f/1.5, 30$^\circ$ off-axis parabola onto the fused silica target at an incidence angle of $\theta=55^\circ$ to a 1.8-\textmu m FWHM spot. With a pulse energy of 2.6~mJ on target, this yields, supposing absence of spatio-temporal distortions, a peak intensity of $1.0\times10^{19}\:$W/cm$^2$, corresponding to a peak normalized vector potential of $a_0=2.0$. A spatially superposed pre-pulse, created by picking off and then recombining $\approx 4$\% of the main pulse through holey mirrors, is focused to a much larger 13~\textmu m FWHM spot in order to generate a homogeneous plasma expanding into vacuum. The plasma density scale length $L$ was controlled by setting a delay of 2~ps between this pre-pulse and the main driving pulse and measured to be $L=\lambda/20$ with spatial-domain interferometry~\cite{bocoum_sdi_2015}. This optimizes the conditions for ROM SHHG emission~\cite{kahaly_direct_2013,haessler_relativisticSHHG_2020} and we will in the following consider the experimentally observed SHHG emission as such although its photon energy range is below the CWE spectral cutoff. 

The emitted harmonic radiation in the specular direction is then recorded, as a function of the driving near-single-cycle laser waveform, using a home-made XUV spectrometer, including a gold-coated flat-field grating (600 lines/mm, 85.3$^\circ$ incidence), which lets the beam freely expand in the vertical dimension and images in the horizontal dimension the source point on the plasma-mirror surface onto a coupled micro-channel-plate (MCP) and phosphor screen detector. The MCP is time-gated for $\approx 250\:$ns synchronously with the laser pulses so as to suppress longer background plasma emission. The phosphor screen is finally imaged by a charged-coupled-device (CCD) camera to record angularly resolved ($[-30,30]\:$mrad) SHHG spectra. 

The CEP of the driving pulses was measured with a home-built f--2f spectral interferometer integrating over 15 laser shots. Since our 3.5-fs laser pulses already have an octave-spanning spectral width this did not include additional spectral broadening which could couple pulse energy fluctuations to phase measurement errors. The laser oscillator was CEP-locked, but the f--2f measurement did not feed back to a slow CEP locking loop. The residual random CEP noise over a 200-ms time window was observed to be $\lesssim500\:$mrad on top of a slow quasi-linear CEP drift of $\approx600\:$mrad/s. 

\begin{figure}[tb]
	\centering
	\includegraphics[width=0.55\textwidth]{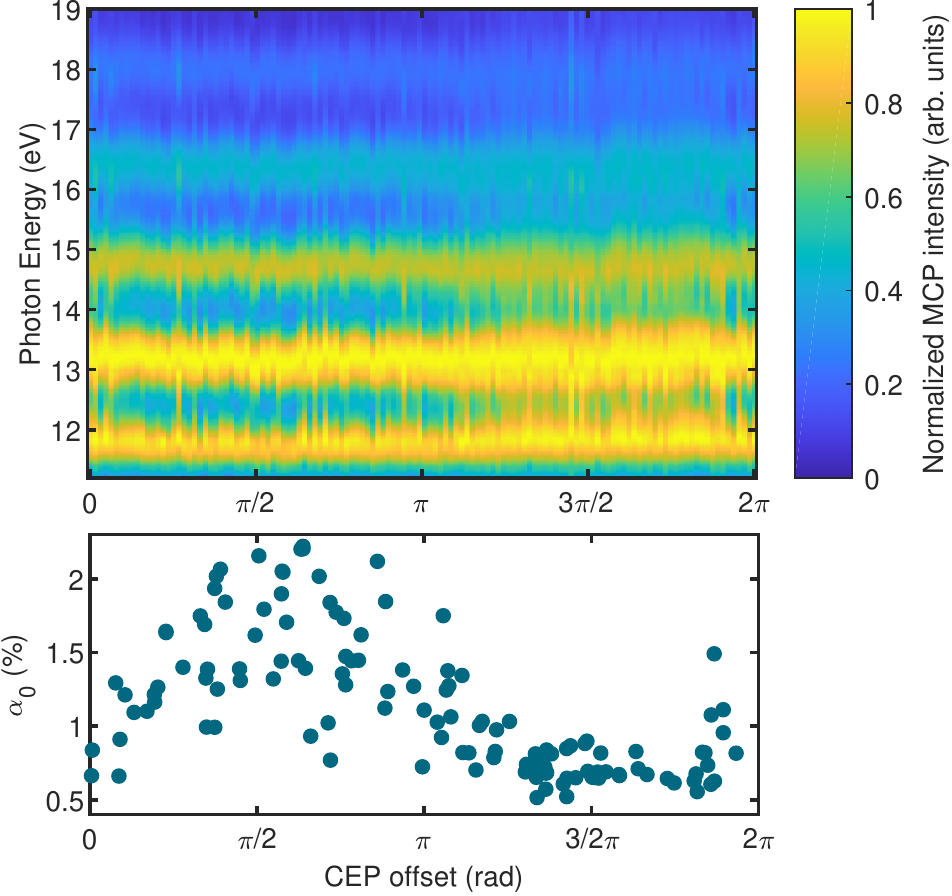}
	\caption{CEP dependence of SHHG spectra, integrated over 30 laser shots. (a) Normalized MCP intensity, angularly integrated over the full detected range \mbox{$[-30,30]\:$mrad}, not corrected for the spectral dependence of the grating and MCP efficiency for better visibility. (b) Relative intensities of satellite pulses in the Fourier-limited temporal intensity corresponding to the SHHG spectra in panel (a).}
	\label{fig:expscan}
\end{figure}

This allowed considering the CEP as sufficiently stable during short bursts of 30 pulses, over which we integrated to record SHHG spectra. A series of such 30-ms long acquisitions were recorded in $\approx4\:$s intervals, so that the CEP is to be considered random from one recorded spectrum to the next. The simultaneously recorded relative CEP is then used to tag the SHHG spectra which yields the CEP-dependence shown in Figure~\ref{fig:expscan}a. Clearly, the harmonic spectral modulation depth is a smooth function of the CEP offset and almost completely disappears for about half of all driving pulse waveforms around a relative CEP offset of $\approx 3\pi/2$. Note that without CEP-tagging, the successively recorded SHHG spectra present random variations, which reassures us in the conclusion that the observed variations are due to the varying CEP of the 3.5-fs driving pulses. Fluctuations of other parameters such as pulse duration and energy are very small ($<2\%$) and cannot account for the observed variability.

\begin{figure}[tb]
	\centering
	\includegraphics[width=0.65\textwidth]{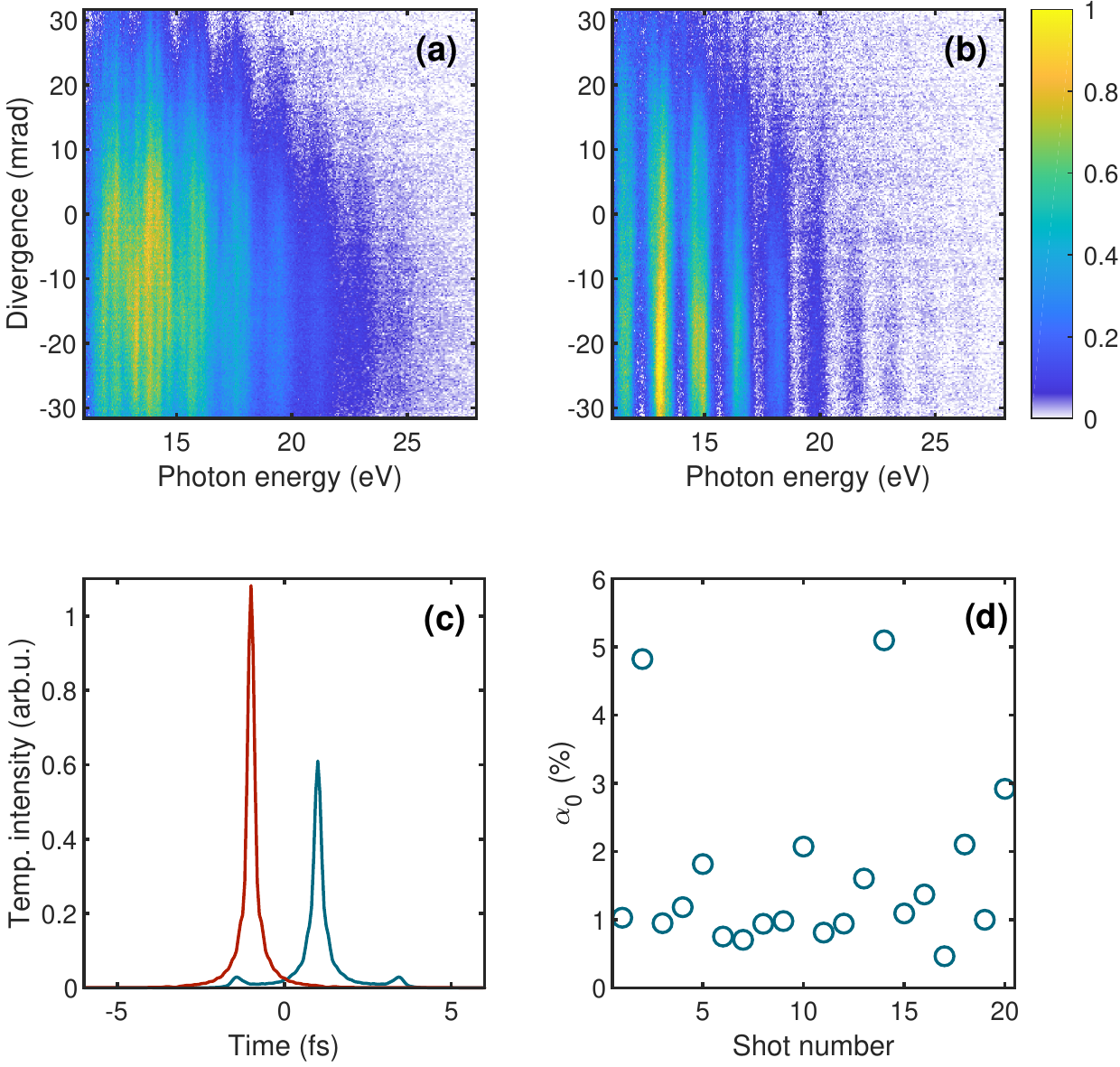}
	\caption{Single shot acquisitions.  Angularly resolved MCP intensity for shots no.~17 (a) and no.~2 (b), not normalized and therefore directly comparable. After angular integration over \mbox{$[-30,0]\:$mrad} and correction for the spectral dependence of the grating and MCP efficiency, Fourier-limited temporal intensity for shots no.~17 (red) and no.~2 (blue). The pulses are delayed by $\pm1\:$fs for better visibility (c). For all 20 single-shot acquisitions, relative intensities $\alpha_0$ of satellite pulses in the Fourier-limited temporal intensity (d). }
	\label{fig:singleshots}
\end{figure}

All our recorded spectra are continua in the sense that the spectrally large harmonics partially overlap. The varying spectral modulation depth suggests, in the time domain, a varying isolation degree of the dominant attosecond pulse relative to neighboring satellite pulses spaced by approximately one laser period of 2.4~fs. In order to quantify this variation, we consider the Fourier-limited temporal intensity profiles corresponding to the measured SHHG spectra (from 11 to 29~eV, corrected for the spectral dependence of the XUV grating efficiency and the MCP response). This necessarily yields symmetric temporal profiles which for our data corerspond to a dominating central attosecond pulse surrounded on either side by two equal weaker satellite pulses (cp. the profiles shown for single shot acquisitions in figure~\ref{fig:singleshots}). 

We plot in figure~\ref{fig:expscan}b the CEP-dependence of the relative intensity $\alpha_0$ of these satellite pulses as compared to the central main pulse. The result confirms our conclusion from the qualitative discussion of the modulation depth visible in figure~\ref{fig:expscan}a: the spectral modulation depth varies smoothly as a function of the driving pulse CEP and is minimal for about half of all CEP values.

The satellite pulse intensities $\alpha_0$ are all so small (between 0.5 and 2\%) that one may want to classify the temporal profiles as isolated attosecond pulses for all CEPs. Of course, this would be a precipitated claim and any statement about a temporal profile deduced only from a spectral intensity requires justified assumptions about the spectral phase. We will discuss this point in section~\ref{sec:discussion}. For now, we consider $\alpha_0$ only as a quantitative measure for the continous nature of the recorded SHHG spectra. 

Since the data in figure~\ref{fig:expscan} are obtained by integrating over 30 consecutive pulses at 1~kHz, any fast jitter of the interaction conditions, notably the CEP, could have washed out the harmonic spectral modulation leading to its underestimation~\cite{goulielmakis_single-cycle_2008}.



We therefore made a series of 20 single-shot acquisitions in $\approx4\:$s intervals, i.e. again for randomly varying CEP. Figures~\ref{fig:singleshots}a,b shows the strikingly different angularly resolved SHHG spectra for two selected shots. We can confirm that even in single-shot acquisitions, we observe spectral continua with almost completely absent harmonic spectral modulation. We also confirm our expectation that the absence of averaging in single-shot acqiusition lets us observe much deeper harmonic spectral modulations for some shots. 

Figures~\ref{fig:singleshots}c,d analyze the harmonic spectra after angular integration over the range \mbox{$[-30,0]\:$mrad}, i.e. over the beam center only. This reduces possible contributions from more divergent CWE harmonics in the harmonic beam periphery (cf. the discussion in section~\ref{sec:discussion}) as well as the possible influence of spatial beam distortions due to aberrations in the harmonic beam imaging. In figures~\ref{fig:singleshots}a,b, it is clearly apparent that the continuous spectrum contains more energy than the modulated one. If the spectral continua indeed correspond to isolated attosecond pulses, then this difference is further amplified by concentrating that energy into a single pulse, which leads to the different peak intensities in the Fourier-limited temporal itensitity profiles shown in figures~\ref{fig:singleshots}c for the same selected shots as in figures~\ref{fig:singleshots}a,b. Figure~\ref{fig:singleshots}d shows the $\alpha_0$ found from the Fourier-limited temporal intensity profiles for all 20 shots. Clearly, the vast majority of shots presents very low spectal modulations : e.g. for 9 out of the 20 pulses $\alpha_0<1$\%, and only for 3 pulses $\alpha_0>2.5$\%. On average over our single-shot acquisitions, we find an $\approx30$\% higher peak intensity of the Fourier-limited temporal profiles for the spectral continua with $\alpha_0<1$\% compared to the modulated spectra with $\alpha_0>2.5$\%.

\section{2D Particle-in-cell simulations}
\label{sec:sim}

In order to interpret our experimental results and support the discussion of the real temporal profiles, we performed a series of 2D particle-in-cell (PIC) simulations using the Warp code~\cite{friedman_computational_2014} with a Yee Maxwell solver. The parameters are chosen to approximately mimic the experimental conditions: a 3.5-fs FWHM Gaussian pulse with central wavelength of $\lambda=800\:$nm and varying CEP is focused to a 1.8-\textmu m FWHM spot to a peak intensity of  $8.5\times10^{18}\:$W/cm$^2$, corresponding to $a_0=2$. The laser impinges at an angle of incidence of 55$^\circ$ onto a plasma with exponential density gradient of scale length $L=\lambda/10$ and a peak plasma density of $400 n_\mathrm{c}$, with the critical density $n_\mathrm{c}$ at 800~nm. 

The simulation box had a size of 60~\textmu m$\times$25~\textmu m with a grid step size of $\lambda/400$ and boundaries absorbing all fields (perfectly matched layer) and particles. 
The simulations used 16 particles per cell and initial fields were generated by constructing them at focus and retropropagating them about one Rayleigh length into the simulation box.  

\begin{figure}[tb]
	\centering
	\includegraphics[width=0.55\textwidth]{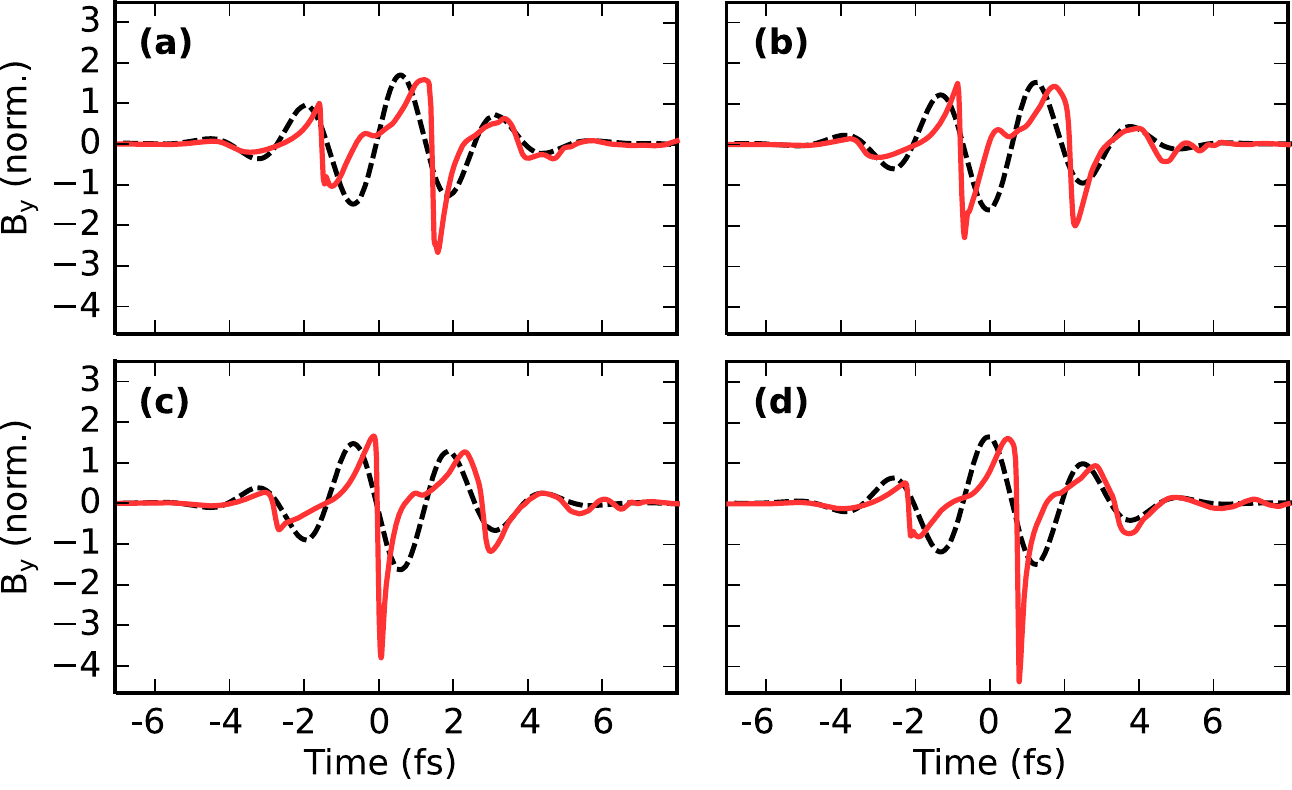} \\ 
	\vspace{10pt}
	\includegraphics[width=0.55\textwidth]{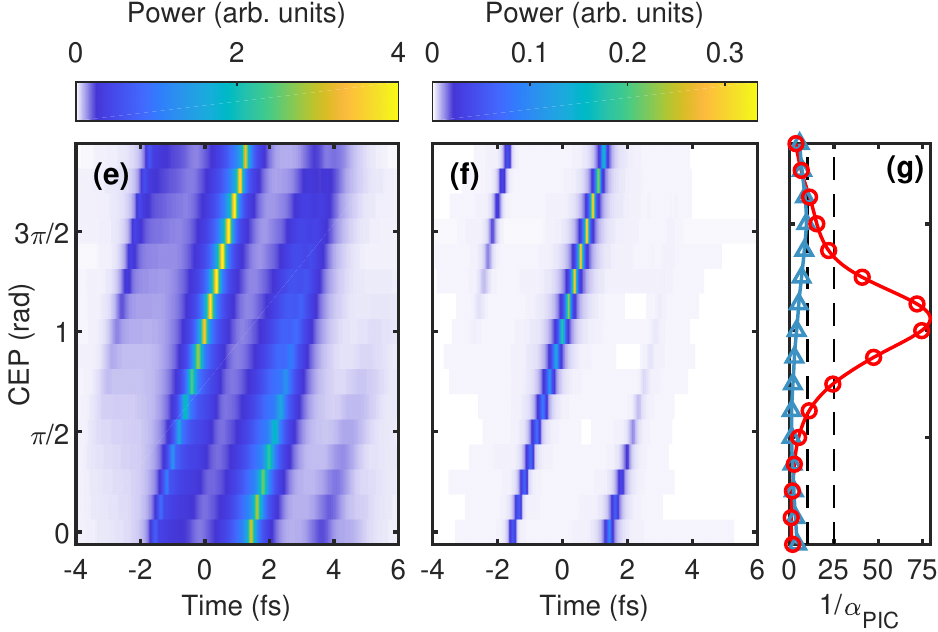}
	\caption{Results of 2D PIC simulations. Magnetic field of the incident (dashed lines, shifted in time by +55~fs) and reflected laser pulse (red lines) for a CEP in focus of 0 (a), $\pi/2$ (b), $\pi$ (c), and $3\pi/2$ (d). Note that these are the CEP values at focus on the plasma mirror surface, whereas the plotted fields are detected about one Rayleigh range away from focus and thus phase shifted due to the Gouy phase slip. Power of the reflected laser pulse, obtained by mutliplying the magnetic field by its Hilbert transform and spatially integrating along the transverse direction, as function if the CEP without any spectral filtering (e) and after selecting photon energies $>10\:$eV only (f). Isolation degree $1/\alpha_\mathrm{PIC}$ (g) for the spectrally filtered (red circles) /unfiltered (blue triangles) attosecond pulse profiles in panels f/e. The dashed line marks $\alpha_\mathrm{PIC} = 10\%$.}
	\label{fig:PIC}
\end{figure}

The dramatic effect of the CEP is evident in the comparison of the incident and reflected laser fields obtained in these simulations shown in figure~\ref{fig:PIC}a-d. The fields are defined such that positive B-fields correspond to the ``push'' phase of the relativistic SHHG process discussed in the introductory section; negative B-fields therefore correspond to the ``pull'' phase. The CEP values considered in figure~\ref{fig:PIC}a-d thus represent different balances of the pushing and pulling laser field cycles: a CEP of $0$ causes the the strongest possible ``push'', CEP$=\pi/2$ leads to two effective field cycles, the first with a strong ``pull'' and the second with a strong ``push'', CEP$=\pi$ causes the strongest ``pull'', and finally CEP$=3\pi/2$ leads to a single effective field cycle with equally strong ``push'' and ``pull''. Clearly, the latter two situations lead to the strongest temporal compression of a single dominant field cycle in the reflected pulse. The enhanced peak fields are compatible with the relativistic electron spring~\cite{gonoskov_theory_2018, thevenet_physics_2016} or coherent synchrotron emission~\cite{anderBruegge2010nanobunching, mikhailova_isolated_2012} models, where SHHG does not only result from a mere phase modulation but from reemission of energy stored in the compressed plasma. 

Multiplying the calculated reflected B-fields by their Hilbert transform yields the pulse intensity, which we spatially integrated along the transverse direction to obtain the power of the reflected pulse. Figure~\ref{fig:PIC}e shows this quantity as a funtion of the incident pulse's CEP. The temporal compression of the field cycles transforms the incident visible 1.5-cycle pulse into an attosecond pulse train even without any spectral filtering. In figure~\ref{fig:PIC}g, we plot the isolation degree, i.e. the inverse of the relative power, $\alpha_\mathrm{PIC}$, of the most intense satellite pulse as compared to that of the central main pulse as function of the CEP. The isolation degree reaches 10 near the CEP$=3\pi/2$, i.e. the resulting reflected pulse could be categorized as an isolated attosecond pulse (with a $1/e$-duration of 0.3~fs). Imposing symmetry around the propagation direction, we can estimate that  this quasi-isolated attosecond pulse contains a 35\% fraction of the incident pulse energy.

Spectral filtering by selecting only photon energies $>10\:$eV lets us consider the same spectral range as in the experiment (cp. figure~\ref{fig:singleshots}). As shown in figure~\ref{fig:PIC}f,g, this transforms the reflected pulses from the PIC simulations into well isolated attosecond pulses with isolation degree $\alpha_\mathrm{PIC}^{-1}>10$ over half the CEP range ($[5\pi/8, 13\pi/8]$), and $\alpha_\mathrm{PIC}^{-1}>25$ over an $1/4$ of the CEP range ($[3\pi/4, 5\pi/4]$), still containing $\approx1\%$ of the incident pulse energy. The ``optimal attosecond pulse''  (highest peak intensity with isolation degree $>25$) is obtained for a CEP=$5\pi/4$ and has a $1/e$-pulse-duration of 0.18~fs.  

Qualitatively, these simulations show that when considering the same spectral range as in our experiment, we expect to generate either a strong single or two weaker attosecond pulses as function of the CEP. A strong satellite pulse ($\alpha_\mathrm{PIC}^{-1}\leq5$) is generated only over $1/3$ of the CEP range ($[-\pi/8, \pi/2]$).

\section{Discussion }
\label{sec:discussion}

Based on the results of the PIC simulations, we will in the following discuss further the possible temporal profile of the SHHG emission generated in our single-shot experiments (cp. fig.~\ref{fig:singleshots}). Qualitatively, we find agreement between the simulation and our experiments in so far as both result in a large majority of driving pulse waveforms leading to a (quasi-)isolated attosecond pulse and spectral continua, and only a narrow CEP-region where two weaker attosecond pulses and therefore strongly modulated spectra are generated (cp. figures~\ref{fig:singleshots}d and the red circles in \ref{fig:PIC}g). 

For a more quantitative comparison, we note that for a given spectral modulation depth, the symmetric Fourier-limited temporal profile is that with the weakest satellite pulse intensities. Considering such a profile with a dominant central pulse sourrounded by two equal satellites with relative intensity $\alpha_0<1$, we find a spectral modulation depth $\nu = [I(\omega)_\mathrm{max} - I(\omega)_\mathrm{min}]/[I(\omega)_\mathrm{max}+I(\omega)_\mathrm{min}] = 4\sqrt{\alpha_0}/(1+4\alpha_0)$, where $I(\omega)$ is the spectral intensity. However, the PIC simulations suggest that the case of a single satellite pulse is much more realistic. Assuming a relative intensity $\alpha_1< 1$ for this satellite pulse, we find $\nu =  2\sqrt{\alpha_1}/(1+\alpha_1)$ -- the same spectral modulation depth thus corresponds to a four times more intense single satellite pulse: $\alpha_1=4\alpha_0$. We could therefore consider an experimentally observed spectral continuum to support a very well isolated attosecond pulse (isolation degree $\alpha_1^{-1}>25$) if the relative satellite pulse intensity in its Fourier-limited temporal intensity profile is $\alpha_0<1\%$. This is the case for 9 out of the 20 single-shot acquisitions considered in figure~\ref{fig:singleshots}, i.e. an even larger fraction that is to be expected from the PIC simulations. 

The strongest spectral modulation we have observed in single-shot acquisitions correspond to $\alpha_0=5\%$, and therefore $\alpha_1^{-1}\approx5$, which correspond to much weaker satellite pulses than those observed in the PIC simulations near CEP=0. This could be due our rather small sample of 20 single-shot acquisitions, which may not contain a driving pulse within the narrow CEP range that leads to such strong satellite pulses. There are however also experimental factors that tend to generally reduce the observed spectral modulation depth even in single shots acquisition. 

\emph{(i)} The undesired detection of incoherent plasma emission polluting the experimental SHHG spectra is strongly reduced by gating the MCP to a 250-ns window around each laser shot, but cannot be complete removed. This limits the achievable HHG-to-background ratio in our experiments to $\sim20$ (observable with longer driver pulses where we would expect the spectral intensities to completely vanish between the harmonic peaks). Therefore the observable spectral modulation depths are $\nu<0.9$, and thus $\alpha_0<10\%$. This can only partly explain the unexpectedly large dominance of very low spectral modulation depths in our experimental spectra.  
 
 \emph{(ii)} Spatio-spectral distortions due to aberrations in the harmonic beam imaging could contribute to washing out spectral modulations.
 
 \emph{(iii)} A non-negligible contribution of the CWE mechanism to the experimentally detected SHHG emission cannot be excluded since we consider photon energies below the its spectral cutoff at $\approx30\:$eV. The CWE-dominated SHHG emission we obtained with the shortest gradient scale length presents strong spatio-spectral distortions, very different from the smooth profiles we are concerned with in this work, obtained with the plasma density gradient scale length adjusted to $L\approx\lambda/20$. This makes us confident that we are examining clearly ROM-dominated SHHG here. 

The importance of both factors \emph{(ii)} and  \emph{(iii)} should increase in the outer parts of the harmonic spatial beam profile, since CWE typically leads to higher divergence~\cite{dromey_diffraction-limited_2009,thaury_high-order_2010}. Indeed, the experimental angle-resolved spectrum in figure~\ref{fig:singleshots}b shows a decreased spectral modulation at higher divergence angles. By selecting only the central 30-mrad-wide divergence cone for the analysis in figures~\ref{fig:singleshots}c,d, we have minimized this influence.

\section{Conclusions}

In conclusion, we have generated high harmonics from plasma mirrors with controlled plasma-density gradient driven by relativistic-intensity 1.5-cycle laser pulses.  We observed a variation of the spectral modulation depth of the SHHG spectra which is clearly correlated to the driving pulses' CEP variations (figure~\ref{fig:expscan}). In single shot-acquisitions, we record XUV spectral continua supporting high-contrast isolated attosecond pulses, with isolation degrees between 10 and 50 for the majority of the driving pulse CEPs. 2D PIC simulations corroborate this interpretation. Our experiments thus represent a significant improvement, in terms of attosecond pulse isolation degree as well as repetition rate, over previous results obtained with 2-cycle drivers at 10-Hz~\cite{jahn_towards_2019,kormin_spectral_2018}, where for the majority of driving pulse CEPs, double or triple attosecond pulse trains where obtained. 

The theoretically predicted percent-level attosecond-pulse generation efficiencies, combined with the possible emission of an isolated attosecond pulse even without any spectral filtering, underline the potential of relativistic plasma-mirrors as extremely-high-brightness secondary sources for future experiments exploring nonlinear interactions in the VUV/XUV range. We must note, however, that we have not measured the XUV flux in our experiments. Experimental laser-to-XUV conversion efficiencies measured for plasma mirror SHHG by other groups are typically $\sim10^{-4}$ in the spectral region beyond 30~eV~\cite{heissler_few-cycle_2012,Roedel2012ultrasteep,yeung_experimental_2017,jahn_towards_2019}, which is about an order of magnitude lower than the predictions of PIC simulations. 

While CEP-tagging is possible even at kHz-repetition rate~\cite{rathje_review_2012, xie_attosecond-recollision-controlled_2012}, it is of course highly desirable to stabilize the driving pulse CEP at its optimal value. We have recently achieved this stabilization with our laser system~\cite{ouille_SN2laser_2020} and are working on harnessing this progress for SHHG experiments.



\ack
This research used resources of the National Energy Research ScientificComputing Center (NERSC). We acknowledge financial support from Agence Nationale pour la Recherche (ANR-11-EQPX-005-ATTOLAB, ANR-14-CE32-0011-03 APERO); Laboratoire d'Excellence Physique: Atomes Lumi\`ere Mati\`ere (LabEx PALM) overseen by the Agence Nationale pour la Recherche as part of the \emph{Investissements d'Avenir} program (ANR-10-LABX-0039); European Research Council (ERC Starting Grant FEMTOELEC 306708); LASERLAB-EUROPE (H2020-EU.1.4.1.2. grant agreement ID 654148), and the R\'egion Ile-de-France (SESAME 2012-ATTOLITE).

\section*{References}
\bibliographystyle{iopart-num}
\bibliography{BIB_towardsIAP}

\end{document}